(a) 2 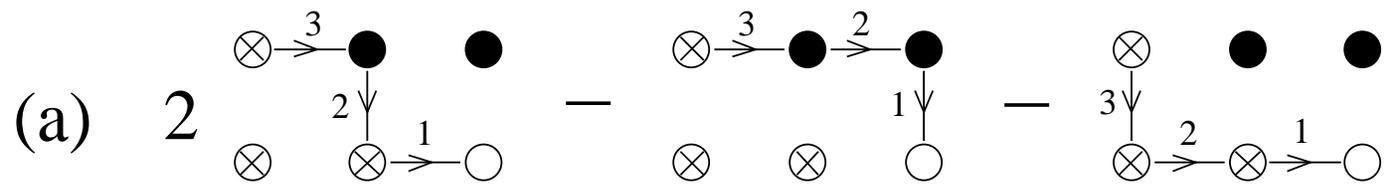

(b) 2 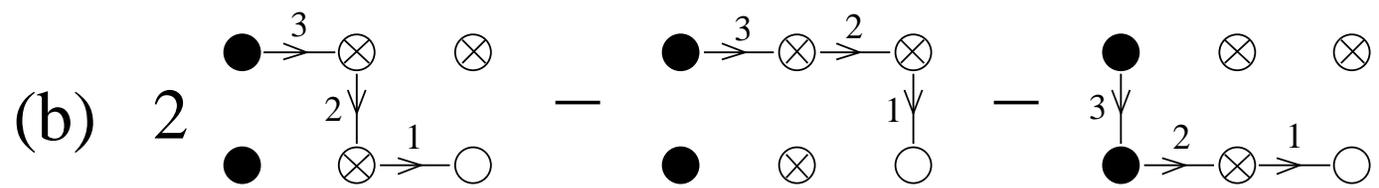

(a) 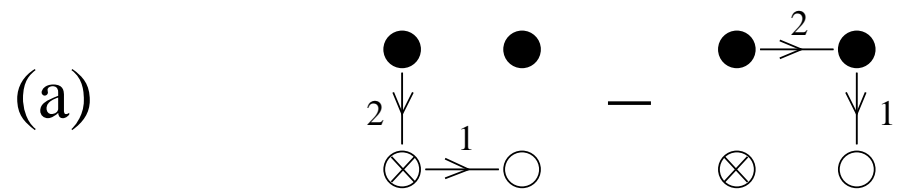

(b) 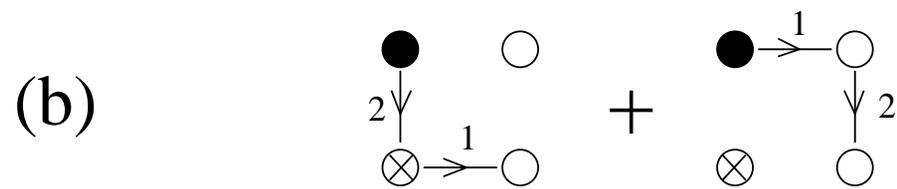

(c) 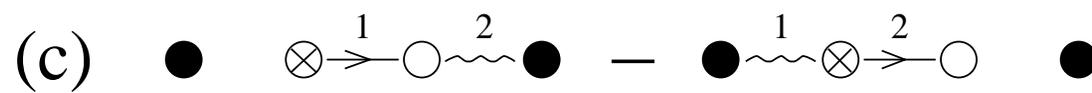

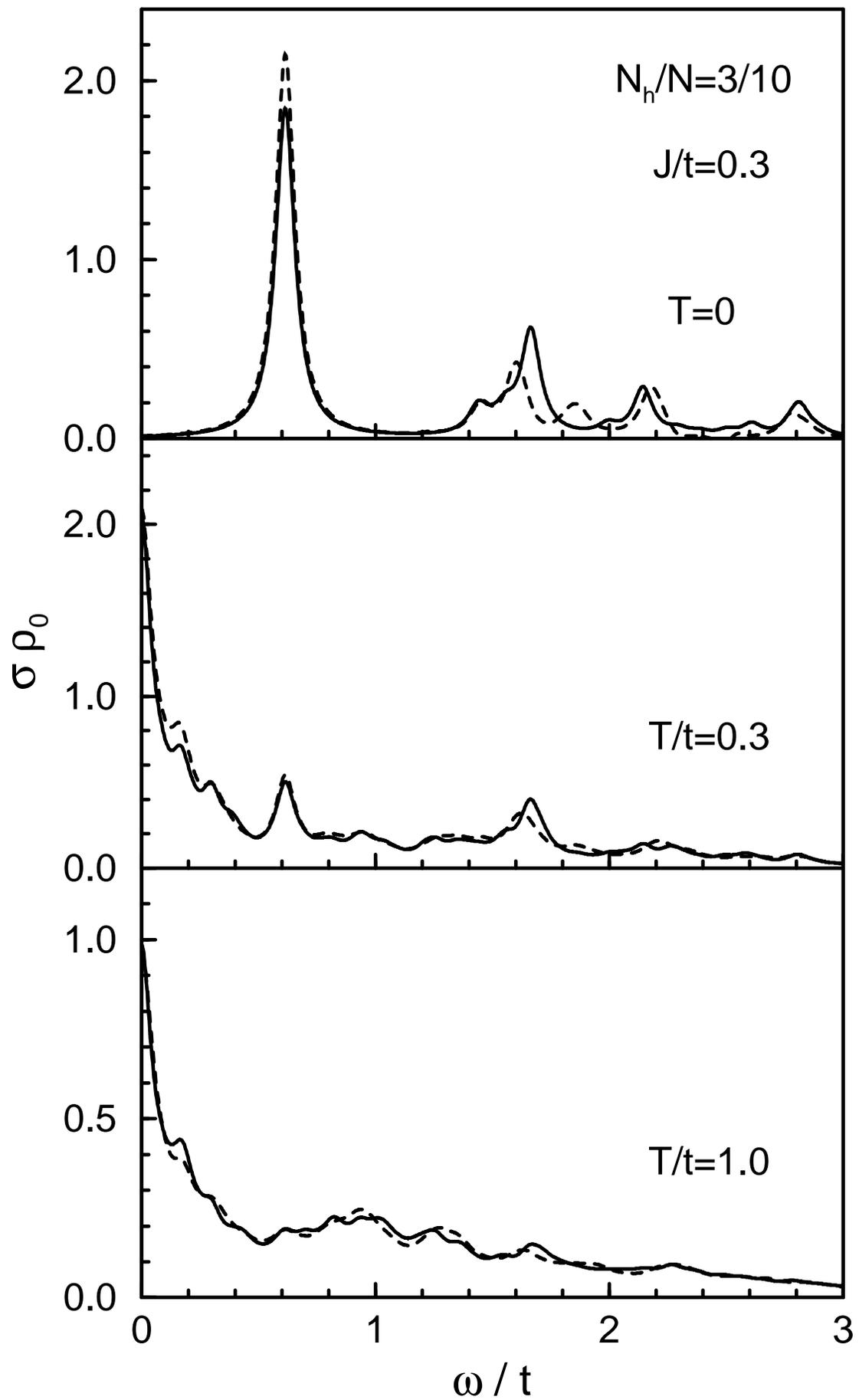

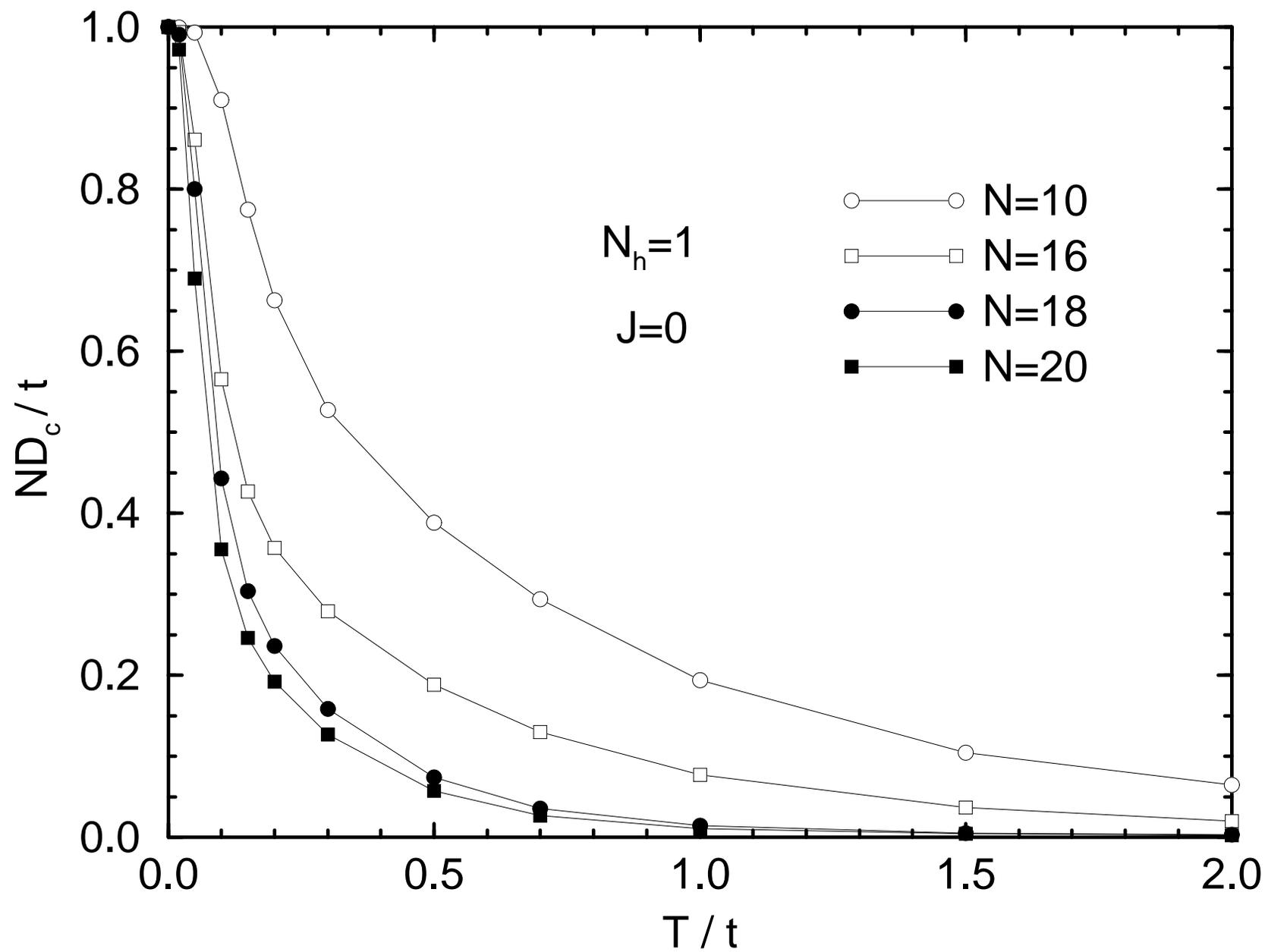

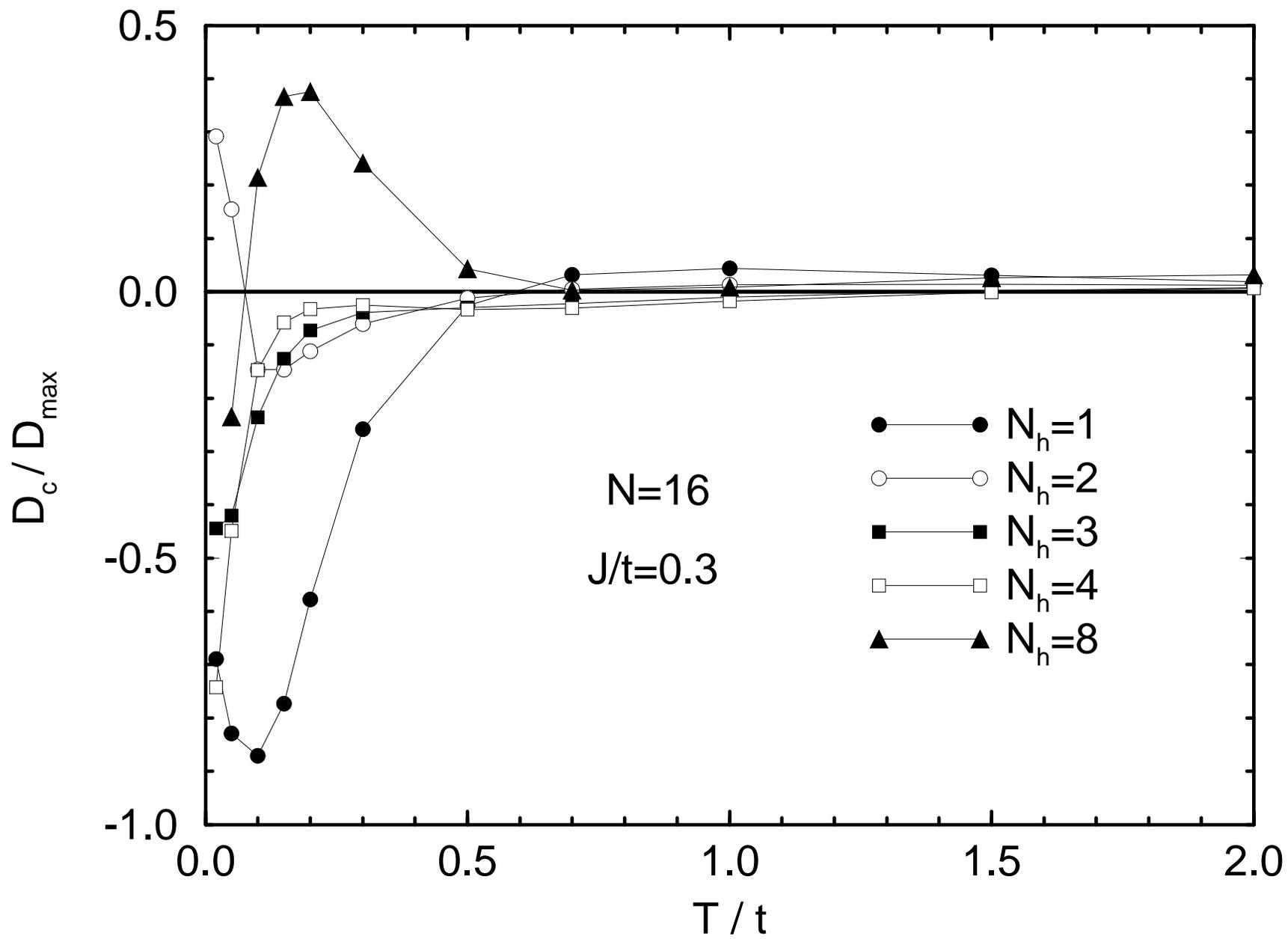

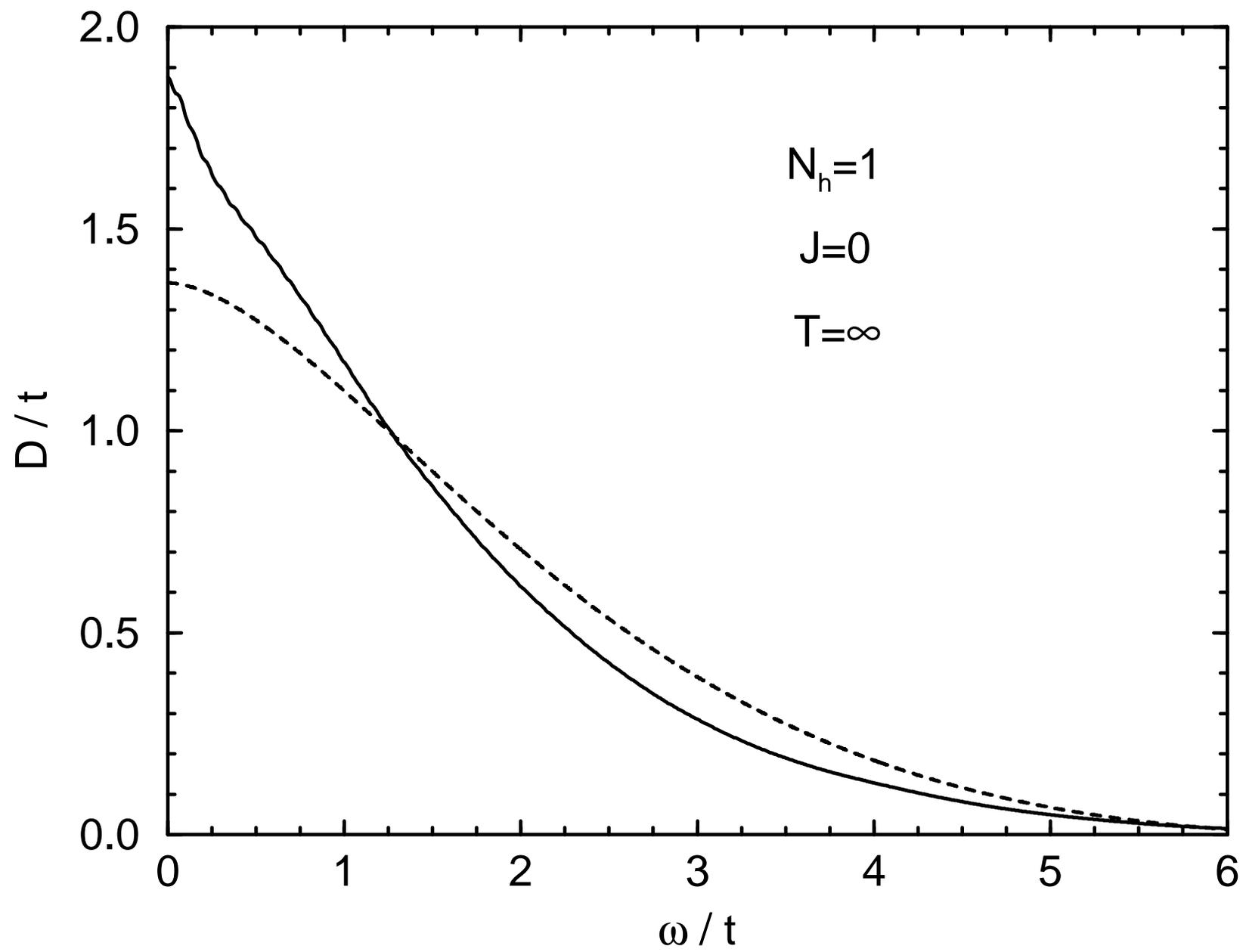

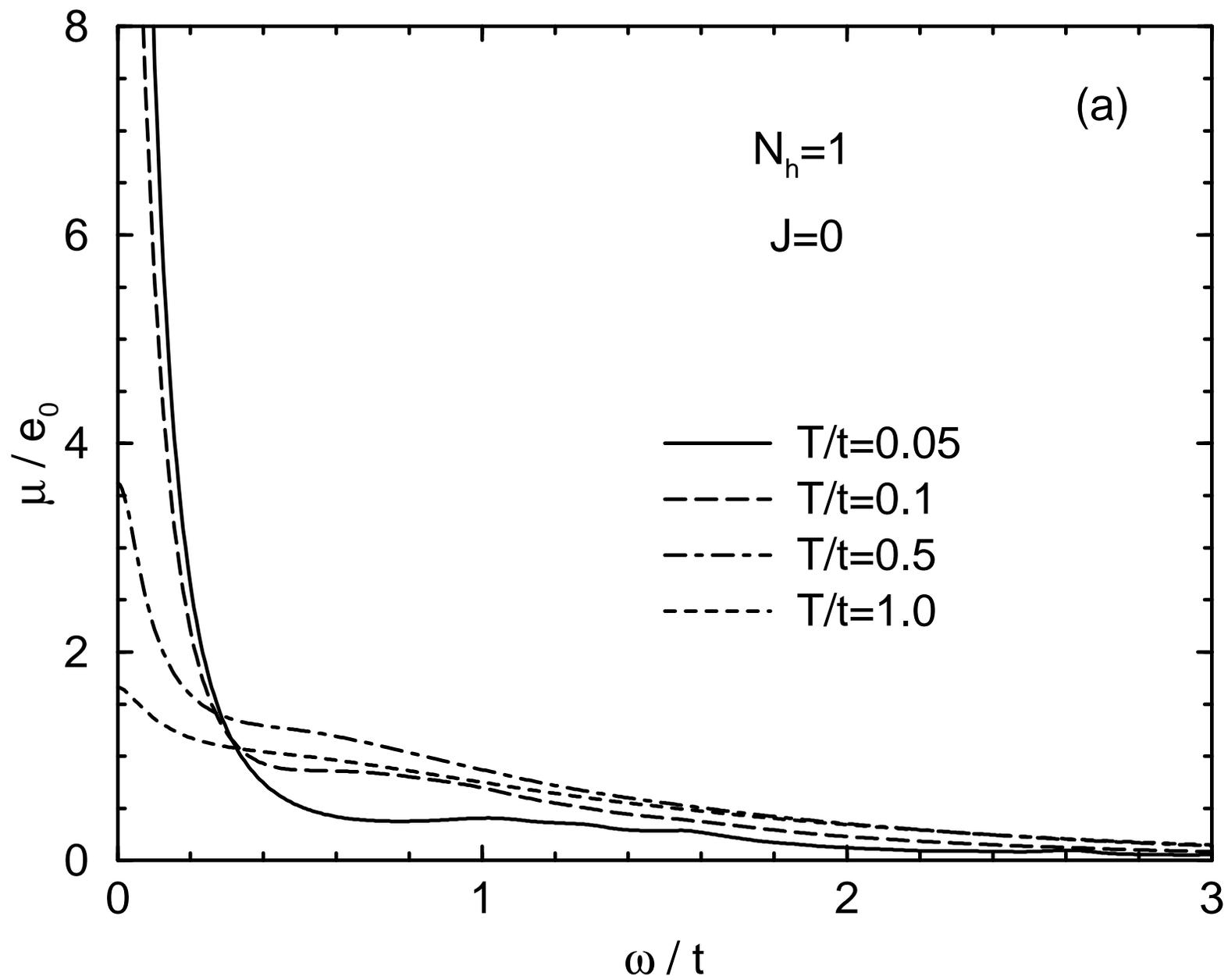

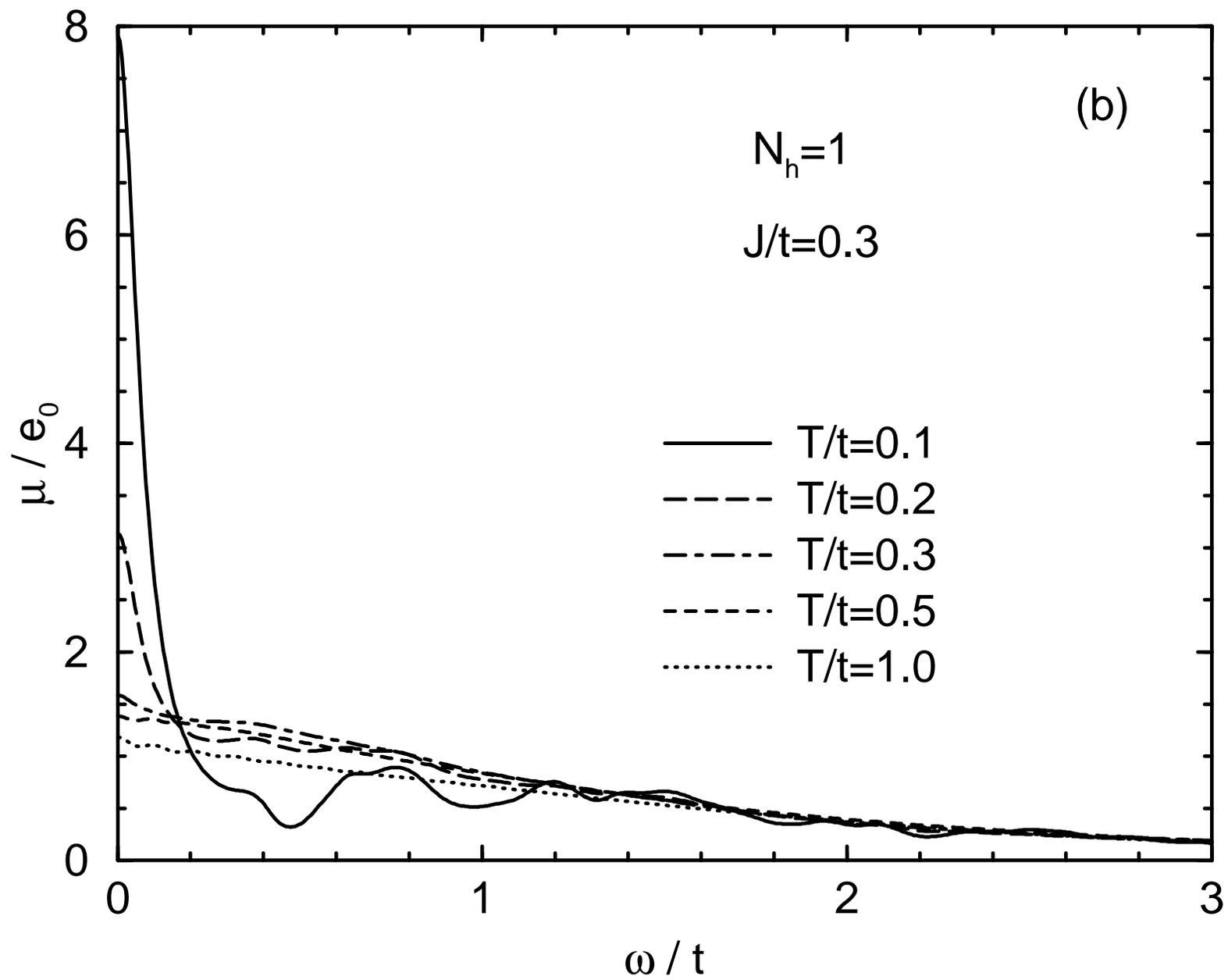

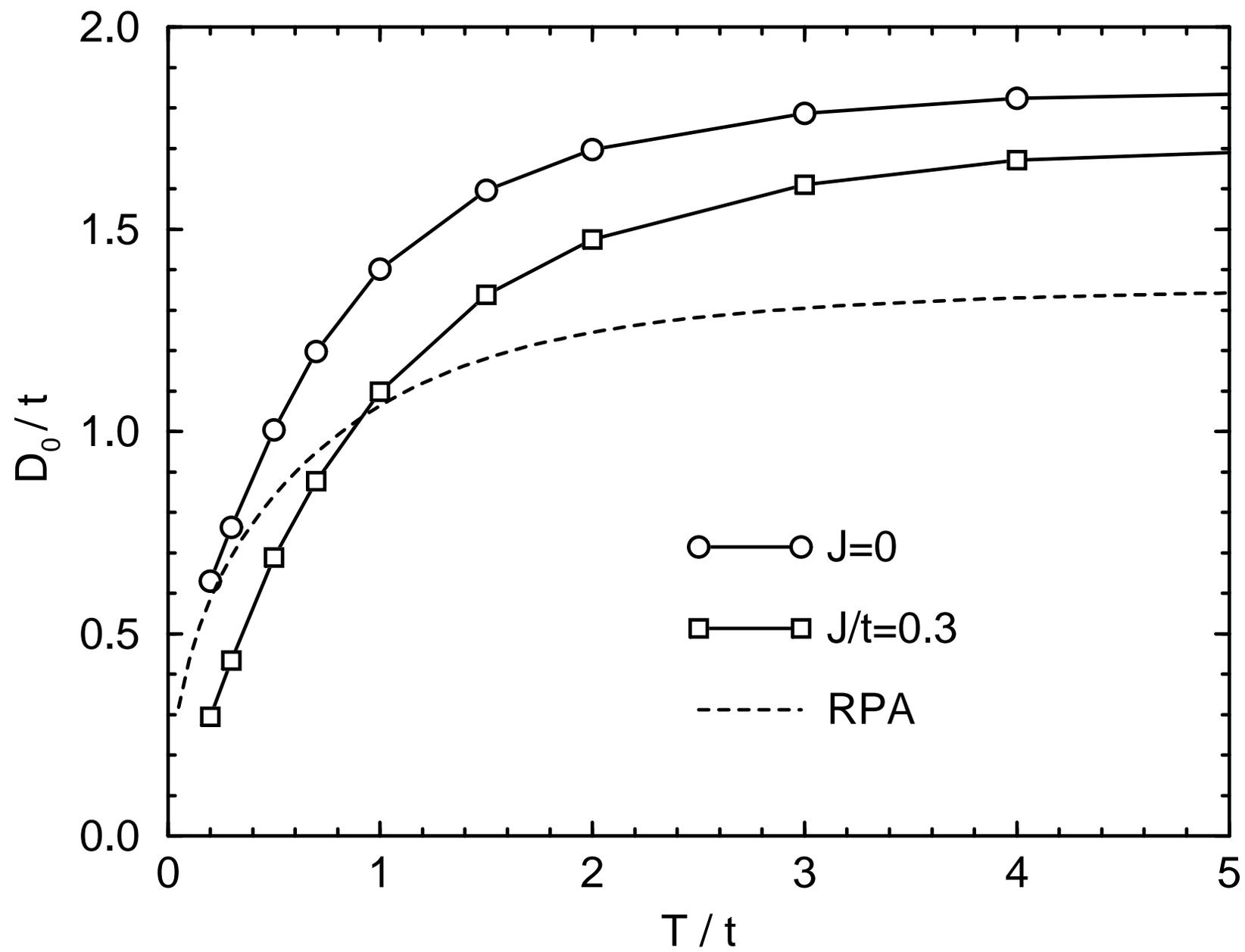

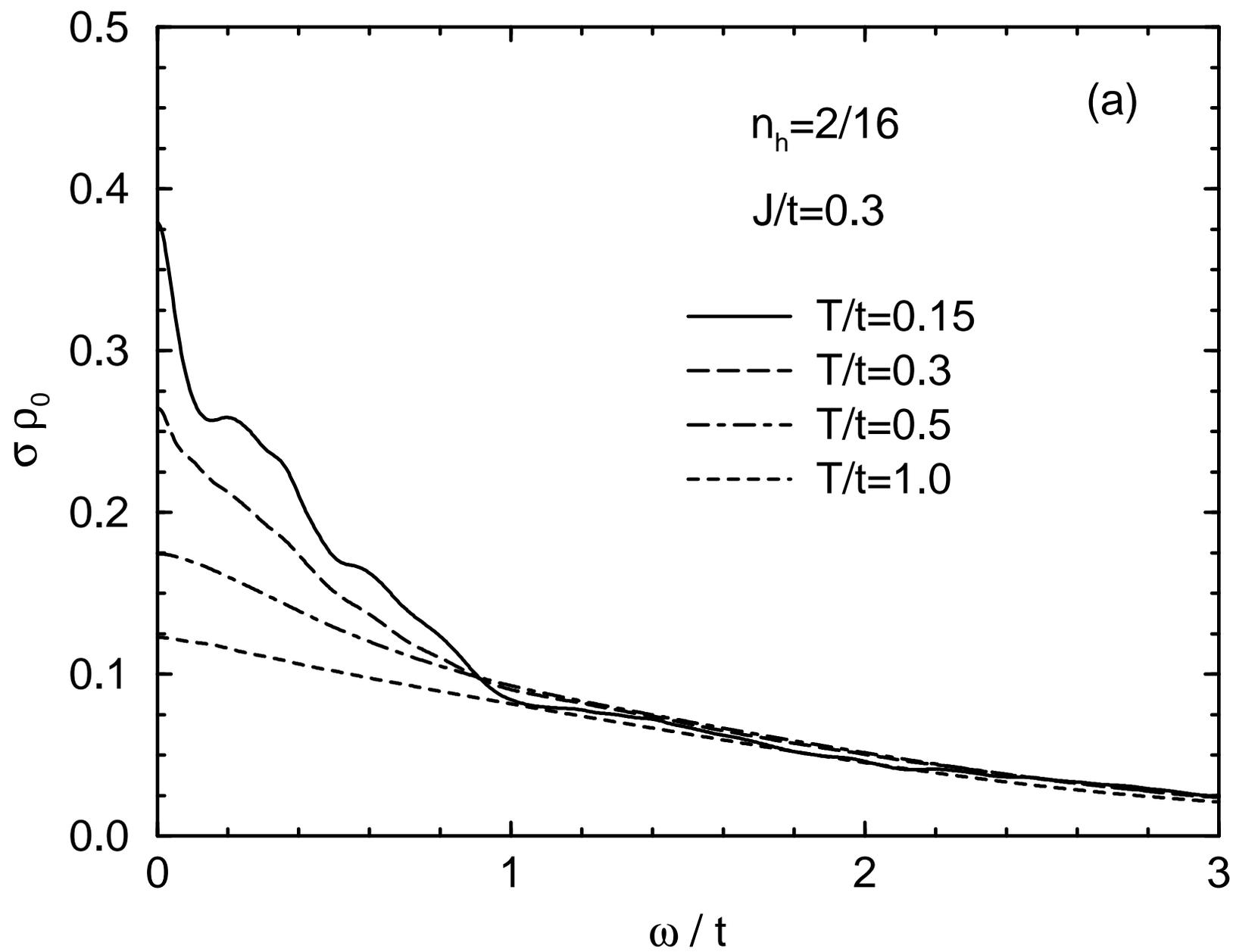

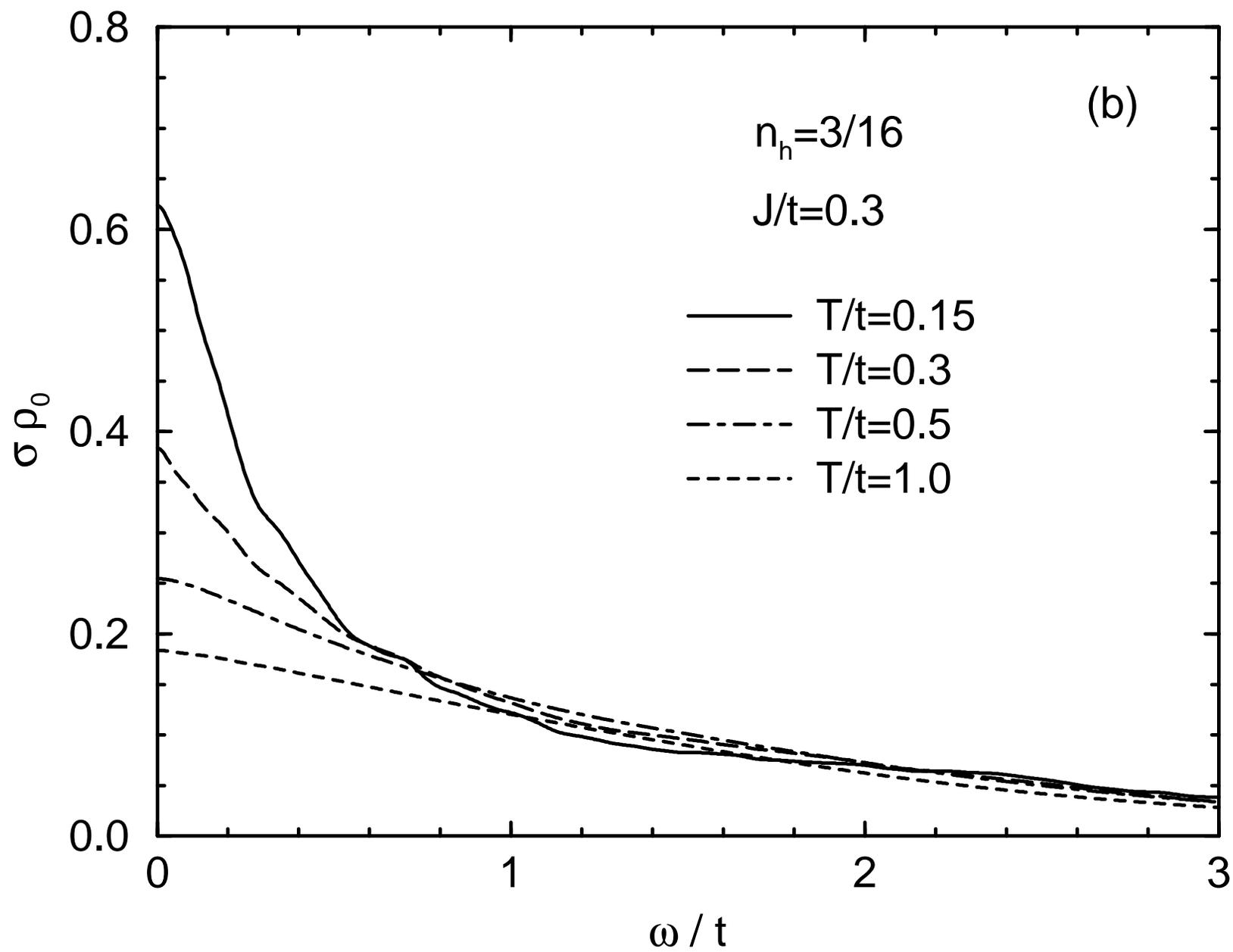

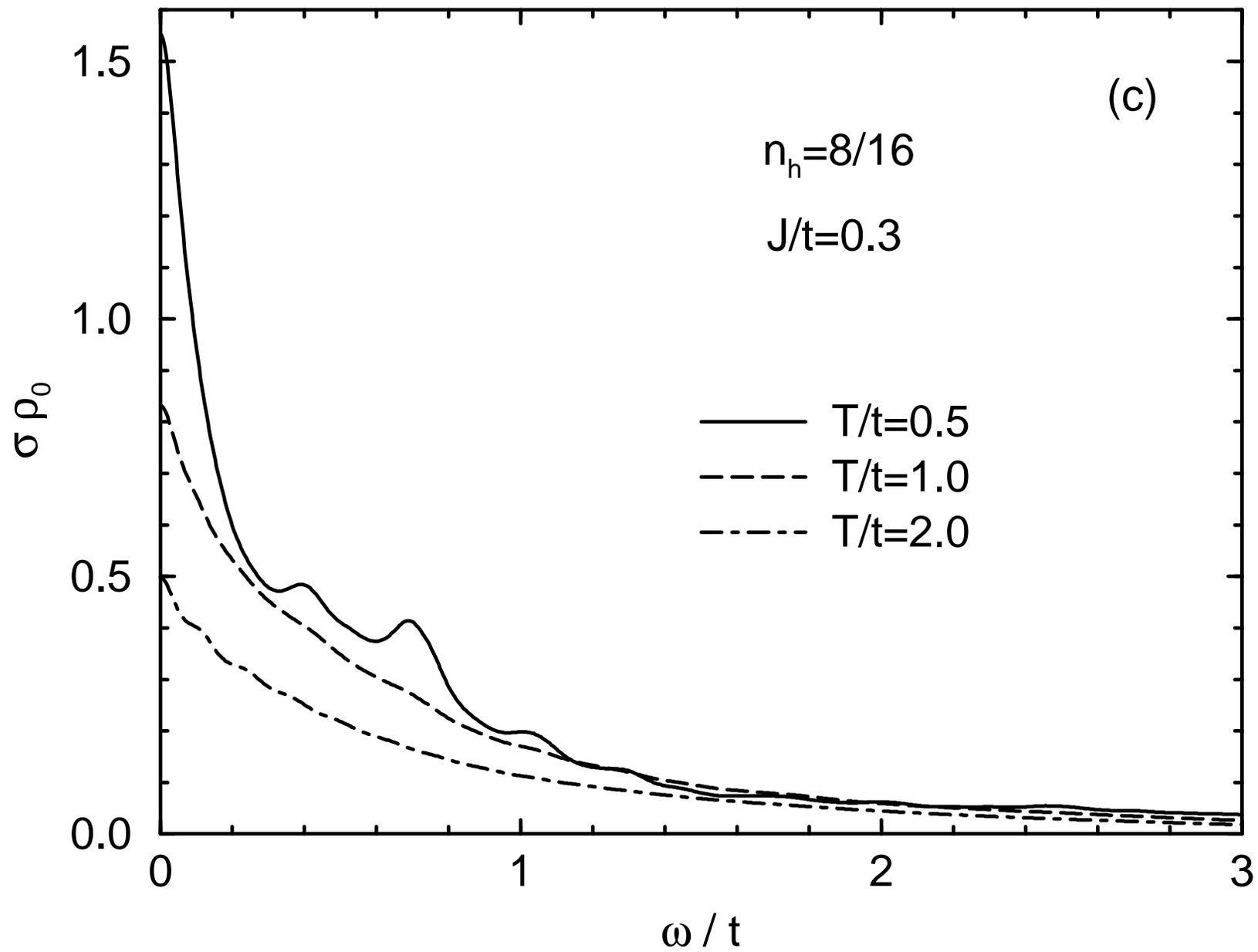

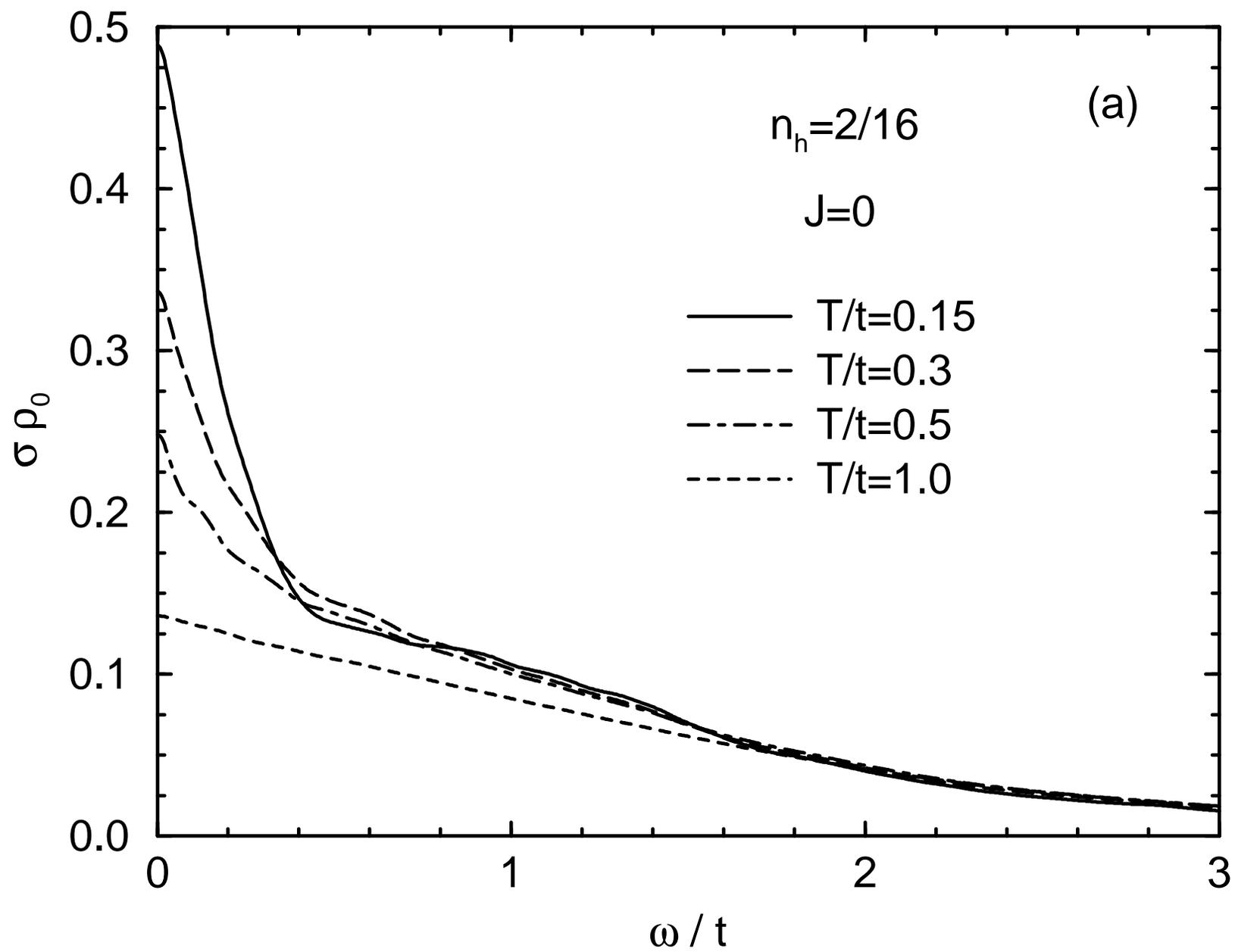

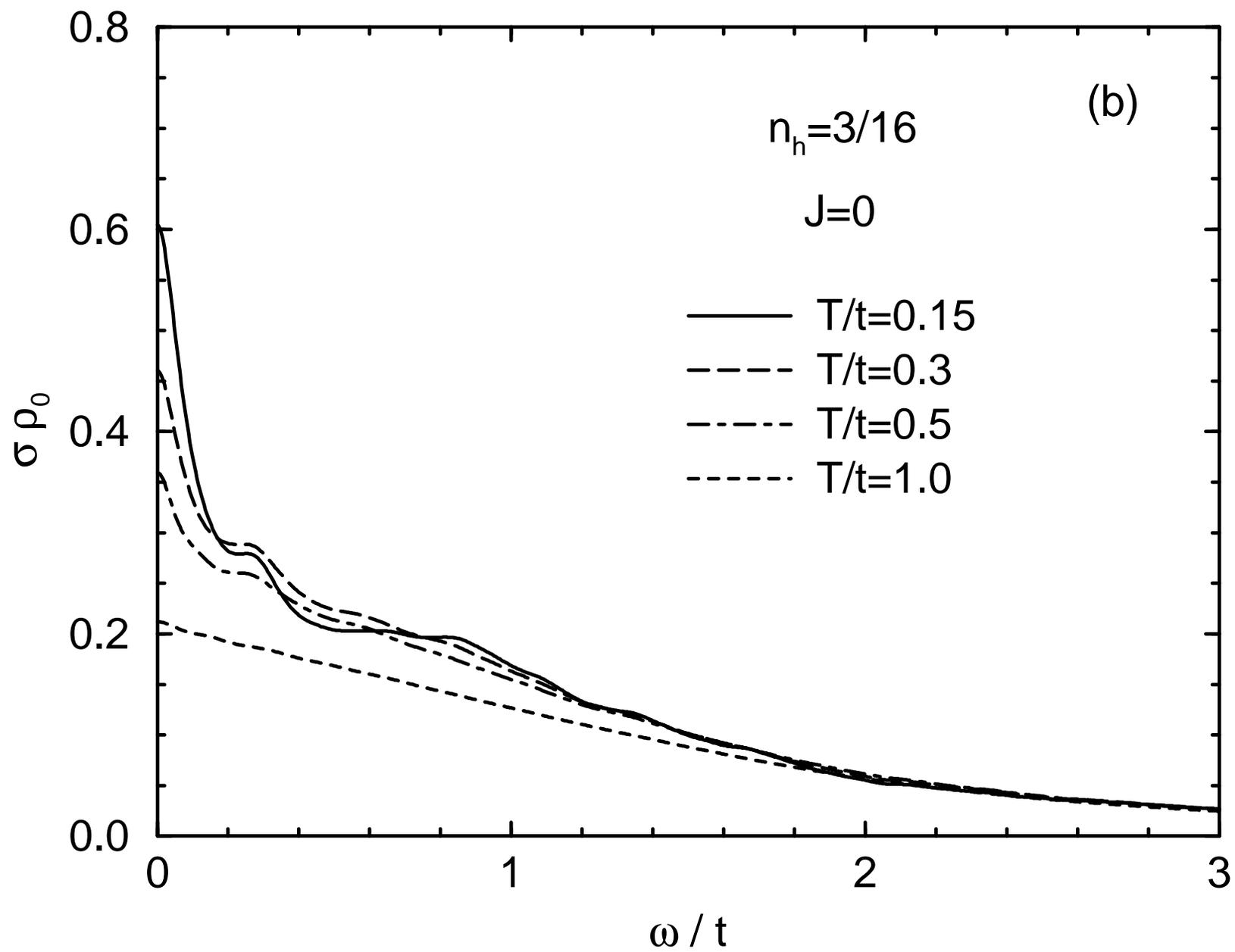

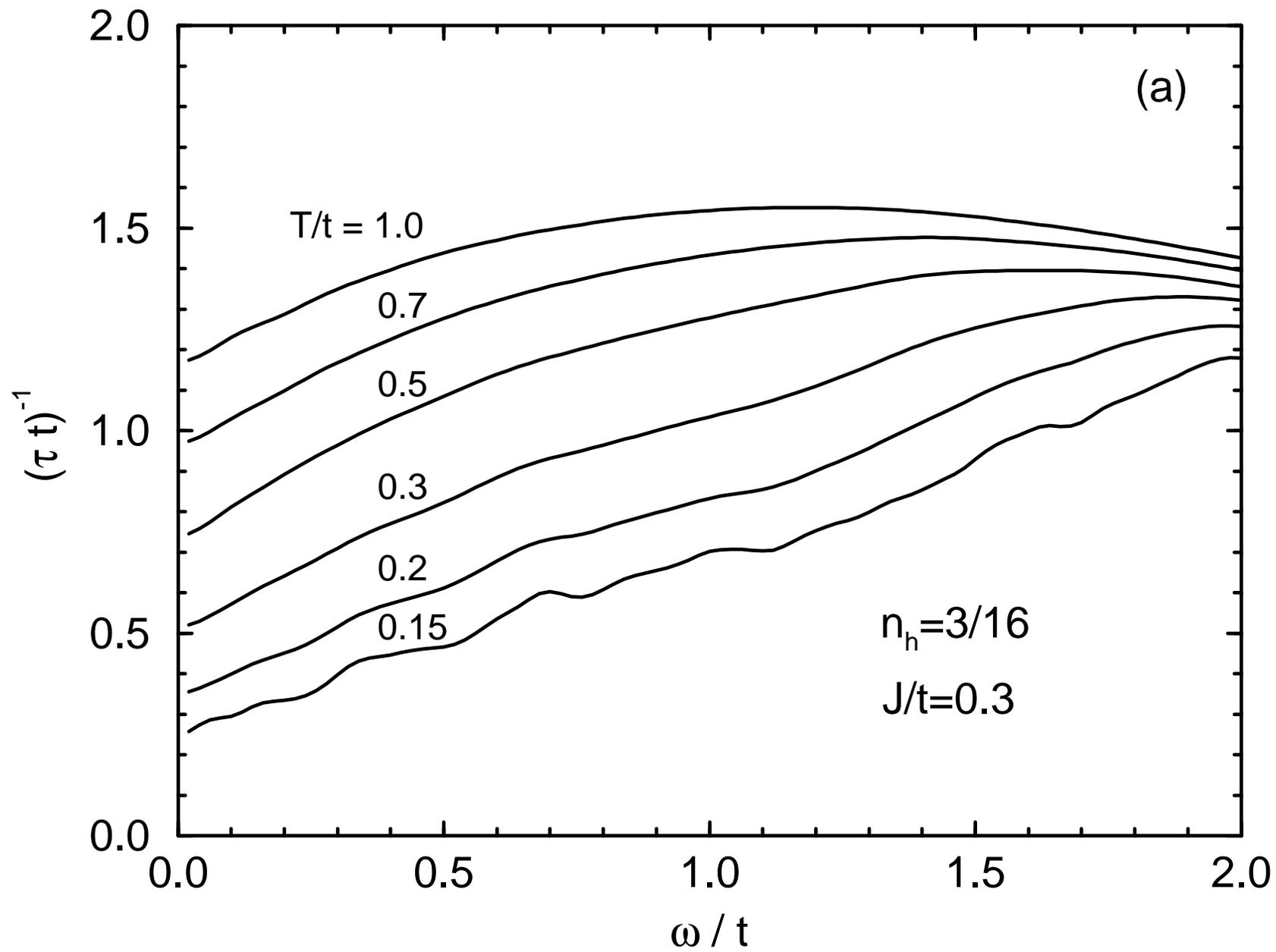

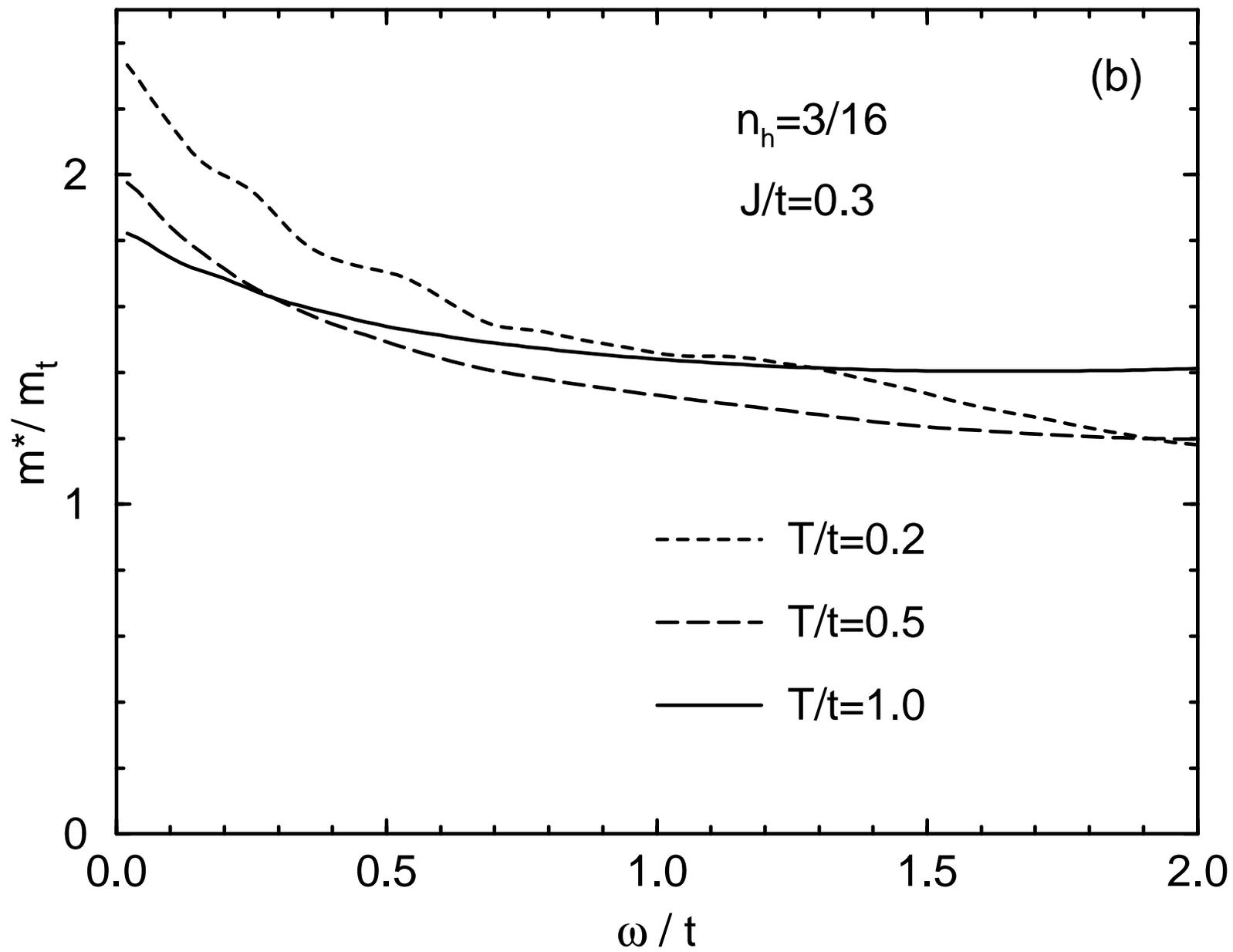

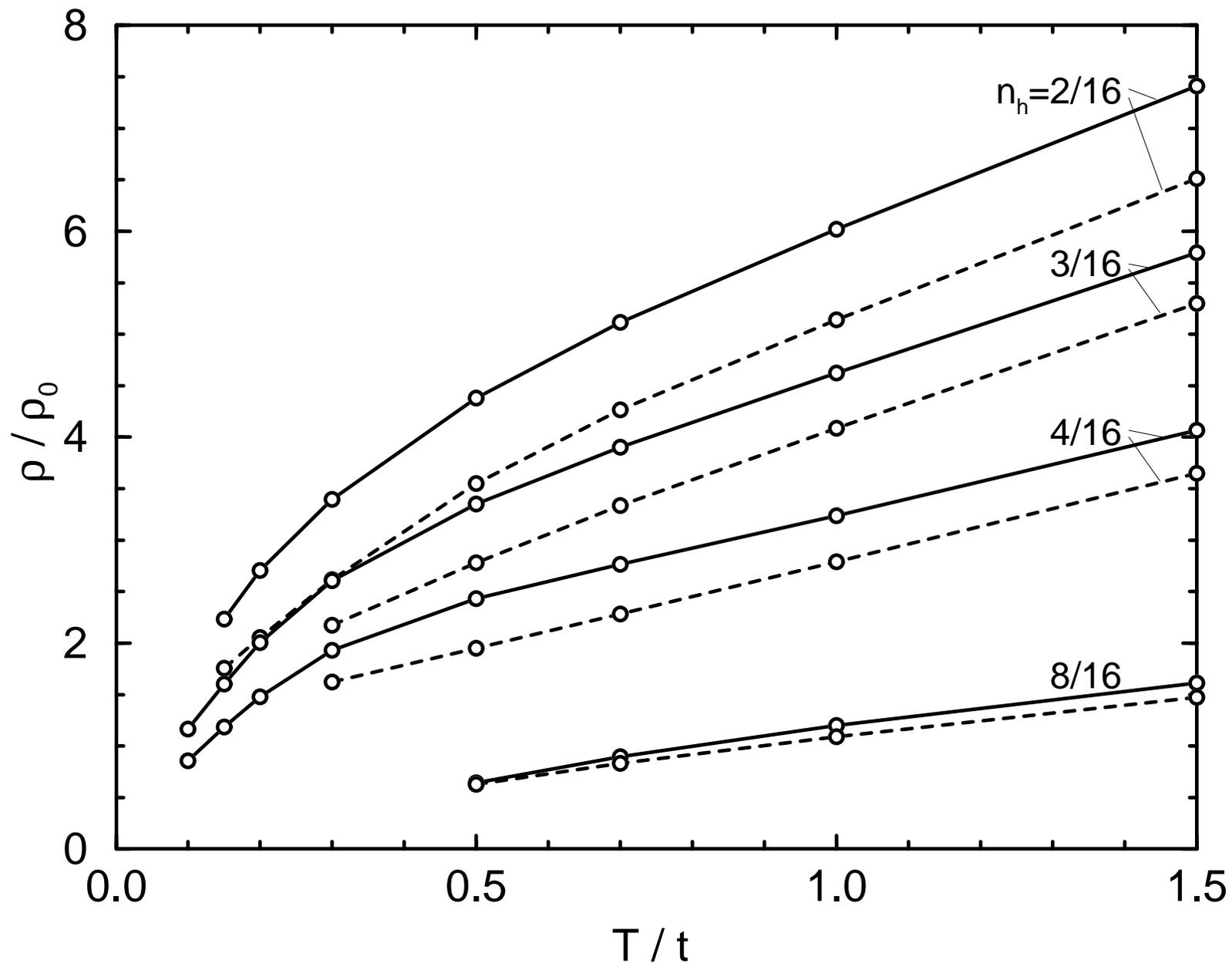

# Charge Dynamics in the Planar $t - J$ Model

J. Jaklič and P. Prelovšek

*J. Stefan Institute, University of Ljubljana, 61111 Ljubljana, Slovenia*

The finite-temperature optical conductivity $\sigma(\omega)$ in the planar $t - J$ model is analysed using recently introduced numerical method based on the Lanczos diagonalization of small systems (up to 20 sites), as well as by analytical approaches, including the method of frequency moments and the retraceable-path approximation. Results for a dynamical mobility of a single hole at elevated temperatures $T > t$ reveal a Gaussian-like $\mu(\omega)$ spectra, however with a nonanalytical behavior at low $\omega$. In the single hole response a difference between the ferromagnetic ($J = 0$) and the antiferromagnetic ($J > 0$) polaron shows up at $T < J$. At larger dopings numerical results in studied systems are consistent with the thermodynamical behavior for $T > T^* \geq 0.1\, t$. $\sigma(\omega)$ spectra show a non-Drude falloff at large frequencies. In particular for 'optimum' doping $n_h \sim 0.2$ we obtain in the low-$\omega, T$ regime the relaxation rate $\tau^{-1} \sim 0.6(\omega + \xi T)$ with $\xi \sim 3$, being consistent with the marginal Fermi liquid concept and experiments. Within the same regime we reproduce the nearly linear variation of dc resistivity $\rho$ with $T$. This behavior is weakly dependent on $J$, provided that $J < t$.

## I. INTRODUCTION

Unusual normal state properties of superconducting cuprates still lack a proper theoretical explanation. There exists however a widespread consensus that most anomalies are closely related to strong electron correlations in these materials. The charge response, as manifested by the optical conductivity $\sigma(\omega)$ and by the dc resistivity $\rho$, has been extensively studied experimentally in cuprates [1,2], as well as theoretically (although with less conclusive results) within some models of strongly correlated systems. This paper is devoted to the study of the finite-temperature conductivity $\sigma(\omega)$ in the prototype planar $t-J$ model and represents an extension of the recently published analysis of the same problem [3].

The $t-J$ model is one of the simplest models for strongly correlated electrons [4],

$$H = -t \sum_{\langle ij \rangle s} (c_{js}^\dagger c_{is} + \text{H.c.}) + J \sum_{\langle ij \rangle} (\vec{S}_i \cdot \vec{S}_j - \frac{1}{4} n_i n_j), \quad (1)$$

where $c_{is}^\dagger (c_{is})$ are projected fermionic operators, taking into account that the double occupancy of sites is not allowed. $n_i$ and $\vec{S}_i$ are corresponding local number and spin operators.

Restricting our discussion to the two-dimensional and higher-dimensional systems, the most conclusive theoretical results have been so far obtained for the particular problem of a single mobile hole introduced into the reference insulator. The insulator is represented by the $t-J$ model at half filling $\bar{n} = 1$, i.e. with all sites singly occupied. In their seminal work Brinkman and Rice [5] solved the problem for the extreme correlation $J = 0$ within the retraceable-path approximation (RPA). They pointed out on an essentially incoherent charge motion and evaluated dc mobility $\mu_0(T)$ of the hole. The latter exhibits a linear variation $\mu_0 \propto T$, at least for high temperatures $T > t$. The same method has been applied, in connection with experimental results on cuprates, to the optical conductivity $\sigma(\omega)$ [6]. Here, the incoherent motion shows up in the slow non-Drude falloff $\sigma(\omega) \propto 1/\omega$ for larger $\omega$. The RPA has been recently justified and applied more rigorously for infinite-D lattices [7]. An alternative approach is the evaluation of frequency moments of $\sigma(\omega)$, starting at $T \gg t$, first applied to the $J = 0$ problem ($U = \infty$ Hubbard model) by Ohata and Kubo [8].

On the other hand, the single-hole ground-state ($T = 0$) optical response has been in recent years well established by numerical studies of small systems via the Lanczos diagonalization method [9–11]. Nevertheless there are important unsolved questions, even for the single hole problem, e.g.: How accurate is the RPA for $\sigma(\omega)$ on the planar lattice? Which are new qualitative spectral features at $T < t$, both for the $J = 0$ and the $J > 0$ case?

More challenging open questions appear at finite hole doping. An evident one is related to the origin of the nearly linear variation $\rho \propto T$ in optimally doped cuprates [2]. Whereas at a low doping level one could possibly treat transport within the (semiconductor) model of independent quasiparticles, i.e. spin polarons, this concept could fail for moderate dopings, e.g. for $n_h > 0.1$, due to the overlap between extended deformations around holes. An attractive proposal in the moderate doping regime is the one of spinons and holons as basic low-energy excitations [12], serving as the starting point for the gauge-theory approaches [13]. On the phenomenological level the marginal Fermi liquid (MFL) concept [14] has been very successful in linking many anomalous normal state properties of cuprates, in particular also $\sigma(\omega)$ and $\rho(T)$, by introducing in the Drude form $\omega$- and $T$-dependent relaxation rate $\tau^{-1}$. Numerical studies of dynamical quantites within the ground state have been so far less conclusive in this regime [11].

The present authors introduced a numerical method [15], allowing for the calculation of dynamical response at $T > 0$. It enables more reliable studies of the normal state (higher-$T$) properties within correlated models, in particular due to a more controlled way of monitoring finite size effects. The method has been applied to the finite-$T$ optical conductivity $\sigma(\omega)$ [3], and more recently also to the spin response [16] in the $t - J$ model. Some important conclusions, in particular those concerning the comparison with experimental results in cuprates, have been given in the short communication [3]. We present here a more complete numerical analysis of the optical and the dc conductivity.

The organization of the paper is as follows: In Sec.II we calculate analytically several frequency moments (up to the fourth order) of $\sigma(\omega)$ at $T \gg t$ within the $t - J$ model on a square lattice. For comparison we also analyse the single hole dynamical mobility within the RPA. Sec.III is devoted to the presentation of the numerical method and related tests. In Sec.IV we first analyse the dynamical and the dc mobility of a single hole, both for $J = 0$ and $J > 0$. Next the results for $\sigma(\omega)$ and $\rho(T)$ at finite hole doping are considered. In particular we extract the phenomenological relaxation time $\tau(\omega, T)$, which we compare to the MFL theory.

## II. ANALYTICAL APPROACHES

### A. Moment method

Within the linear response theory the optical conductivity $\sigma(\omega)$ is given by

$$\sigma(\omega) = \frac{1 - e^{-\beta\omega}}{\omega} C(\omega), \quad C(\omega) = e_0^2 \operatorname{Re} \int_0^\infty d\tau \; e^{i\omega\tau} \langle j(\tau) j \rangle, \tag{2}$$

where $\beta = 1/k_B T$ and $j$ is the particle current density along the chosen axis, e.g. the $x$ axis. In the $t - J$ model (1) with $N$ sites, $j$ can be expressed as (we use in the following units with $\hbar = k_B = 1$ as well as the lattice spacing $a_0 = 1$),

$$j = \frac{it}{\sqrt{N}} \sum_{\langle ij \rangle s} e_{ij} (c_{js}^\dagger c_{is} - \text{H.c.}), \quad e_{ij} = x_j - x_i. \tag{3}$$

The method of frequency moments has been one of the standard ones to analyse the dynamical response [17]. It has been used for the conduction in a correlated system first by Ohata and Kubo [8], where $\sigma(\omega)$ in a simple cubic lattice has been analysed. Analogous approach, combined with the high-$T$ expansion, has been recently used to evaluate other dynamical quantities in the $t - J$ model [18]. We are using in the following the method for the $t - J$ model on a square lattice and restrict our analysis to *high temperatures* $T \gg t$.

The goal is to calculate frequency moments of $\sigma(\omega)$,

$$m_{2n} = \int_{-\infty}^\infty d\omega \; \omega^{2n} \sigma(\omega) = \pi \beta e_0^2 \alpha_{2n}, \tag{4}$$

taking into account that odd moments vanish. For $T \gg t$ moments are related to the short-time expansion of the current-current correlation function, i.e.

$$\langle j(\tau)j \rangle = \sum_{n=0}^{\infty} \frac{(-1)^n}{(2n)!} \alpha_{2n} \tau^{2n} ,$$
$$\alpha_{2n} = \mathrm{Tr}(j^{(n)} j^{(n)})/\mathrm{Tr}\, 1, \qquad j^{(n)} = \mathcal{H}^n j, \qquad (5)$$

and $\mathcal{H}A = [H, A]$.

It is instructive to obtain analytical expressions for few lowest moments $\alpha_{2n}$, since they provide an insight into the processes leading to the current relaxation. The evaluation of the moments is straightforward, but very tedious for larger $n$. We follow a procedure analogous to the one used in [8], summing (tracing) in Eq. (5) over all initial configurations and over processes which restore these configurations.

*The zeroth order moment* is independent of the lattice dimensionality (in hypercubic lattices) [8]. It requires a simple hopping process, an occupied site and an unoccupied site being neighbors,

$$\alpha_0 = \mathrm{Tr}(jj)/\mathrm{Tr}\, 1 = 2t^2 n_e n_h, \qquad (6)$$

where we introduce $n_e = N_e/N$ and $n_h = 1 - n_e = N_h/N$ as the electron and the hole density (per site), respectively.

*The second order moment* involves contributions from the hopping and the spin part in Eq. (1). Let us first consider the hopping contribution, where we can represent both $H_t$ and $j$ as a sum over hoppings between directed pairs of sites,

$$H_t = \sum_{(ij)} H_{ij}, \qquad j = i \sum_{(ij)} e_{ij} H_{ij}, \qquad (7)$$

and

$$j_t^{(1)} = [H_t, j] = H_t j - j H_t . \qquad (8)$$

It is important to realize that due to Eq. (7) $j_t^{(1)}$ vanishes identically in the 1D chain of correlated electrons [5], as well as for hoppings along one direction only, being a manifestation of the current conservation in 1D. On a square lattice nonvanishing contributions to $j_t^{(1)}$ come from the processes on a plaquette, as represented in Figs. 1(a),(b), under the condition that fermionic spins on the plaquette are not equal. Summing over different incoherent contributions to Eq. (5) within a given configuration and different realizations of the plaquette, we obtain

$$\alpha_{2t} = 6t^4 n_e^3 n_h + 8t^4 n_e^2 n_h^2, \qquad (9)$$

where the first and the second term come from the configurations as shown on Fig. 1(a) and Fig. 1(b), respectively. We treat the exchange term $H_J$ in an analogous way and one contributing process to $j_J^{(1)}$ is shown in Fig. 1(c). The result is

$$\alpha_{2J} = 4t^2 J^2 \sum_{l \text{ n.n. } i} \langle (\vec{S}_i \cdot \vec{S}_l)^2 \rangle = \frac{9}{4} t^2 J^2 n_e^2 n_h \ . \qquad (10)$$

*The fourth order moment* requires quite tedious calculations, hence we restrict ourselves to particular cases: either the vicinity of half-filling $n_h \ll 1$ or to nearly empty band $n_e \ll 1$, both in the strong correlation limit $J = 0$. We have to consider the action of the operator

$$j_t^{(2)} = [H_t, j_t^{(1)}] = H_t H_t j - 2 H_t j H_t + j H_t H_t \ . \qquad (11)$$

Again, hopping contributions along only one chosen axis vanish due to $j_t^{(1)} = 0$, leading to $\alpha_4 = 0$ in 1D. The main task is to group all processes which could lead to the same intermediate state through the operation of $j^{(2)}$ and thus contribute coherently to $\alpha_4$. Two examples for $n_h \ll 1$ are in Figs. 2(a),(b). The coherence factor depends on the particular spin configuration. E.g., whereas in Fig. 2(a) three different final states are obtained (resulting in the coherence factor $C = 6$), in Fig. 2(b) only two final states are different, leading to a partial cancellation and $C = 2$. The enumeration of all nonequivalent diagrams (analogous to those in Figs. 2(a),(b)) leads to

$$\alpha_4 = 66 t^6 n_h \ , \qquad J = 0, \qquad n_h \ll 1. \qquad (12)$$

An analogous consideration of processes involving configurations with two fermions (with opposite spins analogous to Fig. 1(b)) on a rectangular plaquette yields

$$\alpha_4 = 48 t^6 n_e^2 \ , \qquad J = 0, \qquad n_e \ll 1. \qquad (13)$$

What can one learn from moments, as presented above for $T \gg t$ ? At *low hole doping* $n_h \ll 1$ one can interpret conductivity in terms of the dynamical mobility $\mu(\omega)$ or diffusion response $D(\omega)$ of independent holes, related by

$$\sigma(\omega) = e_0 n_h \mu(\omega), \qquad \mu(\omega) = \beta e_0 D(\omega), \tag{14}$$

where the latter relation is meaningful only for $T \gg t$. The diffusive motion of holes in 2D (in contrast to the nondiffusive one in 1D) emerges due to the noncommuting processes around individual plaquettes, as presented in Figs. 1,2. The lowest frequency moments of $D(\omega)$ for $n_h \ll 1$ follow from Eqs. (6,9,10,12),

$$\langle \omega^2 \rangle = \frac{\alpha_2}{\alpha_0} = 3t^2 + \frac{9}{16} J^2,$$
$$\langle \omega^4 \rangle = \frac{\alpha_4}{\alpha_0} = 33 t^4, \qquad J = 0 \ . \tag{15}$$

From Eq. (15) we note that the exchange term contributes less to the hole scattering (provided that $J < t$) than the strong correlation requirement of no double occupancy. It is also interesting that lowest moments predict a nearly Gaussian form of $D(\omega)$, since $\langle \omega^4 \rangle \sim 3 \langle \omega^2 \rangle^2$. One can therefore extract an approximative dc diffusion constant $D_0$ as

$$D_0 \sim \frac{\pi \alpha_0}{n_h \sqrt{2\pi \langle \omega^2 \rangle}} = \sqrt{\frac{2\pi}{3}} t = 1.447 t \ , \tag{16}$$

to be compared later with other methods.

*At low electron density* $n_e \ll 1$ the current scattering is essentially different. The frequency moments

$$\langle \omega^2 \rangle = 4 t^2 n_e, \qquad \langle \omega^4 \rangle = 24 t^4 n_e \tag{17}$$

are non-Gaussian $\langle \omega^4 \rangle \gg 3 \langle \omega^2 \rangle^2$. It is easy to recognize that processes included in Eqs. (9,13) contain only two-body collisions of electrons with opposite spins. For configurations with a pair of electrons with parallel spins the current can decay only due to the processes involving other electrons, hence terms with $\alpha_4 \propto n_e^3$ are crucial etc. Therefore, one can expect quite different spectra $\sigma(\omega)$ with possibly more singular dependence at $\omega \ll t$.

## B. Retraceable path approximation

The RPA has been introduced [5] for the analysis of a single-hole motion in the strongly correlated system with $J = 0$. Assuming that the motion of a hole in the spin background is entirely incoherent, i.e. without closed loops, only retraced paths contribute to the local propagator $G(\omega)$, approximated as

$$G(\omega) = \frac{2(z-1)}{(z-2)\omega + z\sqrt{\omega^2 - \omega_0^2}}, \tag{18}$$

where $z$ is the coordination number of a lattice (on a square lattice $z = 4$) and $\omega_0 = 2\sqrt{z-1}\,t$. The RPA is exact in 1D, on Bethe lattices, as well as on infinite-D lattices [7]. Within the RPA it is possible [5,6] also to evaluate the optical conductivity $\sigma(\omega)$. The mobility, defined by Eq. (14), can be expressed as

$$\mu(\omega) = 2t^2 \frac{1 - e^{-\beta\omega}}{\omega} \int_{-\omega_0}^{\omega_0} d\omega' e^{-\beta\omega'} \mathcal{F}(\omega', \omega + \omega') \Big/ \int_{-\omega_0}^{\omega_0} d\omega' e^{-\beta\omega'} G''(\omega'), \tag{19}$$

where

$$\mathcal{F}(\omega_1, \omega_2) = \sum_{s_1, s_2 = \pm 1} s_1 s_2 F(\omega_1 + is_1\delta, \omega_2 + is_2\delta),$$
$$F(\omega_1, \omega_2) = G(\omega_1)G(\omega_2)\left[1 + \frac{1}{\alpha_1\alpha_2} - \frac{\alpha_1^2 + \alpha_2^2 - 2}{\alpha_1\alpha_2(\alpha_1\alpha_2 - 1)}\right], \tag{20}$$
$$\alpha_{1,2} = \frac{1}{2t}\left[\omega_{1,2} + \sqrt{\omega_{1,2}^2 - \omega_0^2}\right].$$

For $T \gg t$ the diffusivity spectra $D(\omega) = T\mu(\omega)/e_0$, resulting from Eqs. (18-20), are presented together with numerical result in Fig. 6. In spite of the finite cutoff at $\omega_{max} = 2\omega_0 = \sqrt{12}\,t$, spectra have a close resemblance to the Gaussian function. This is clear from the lowest moments within the RPA,

$$\langle \omega^2 \rangle = 4t^2, \qquad \langle \omega^4 \rangle = 48t^4, \tag{21}$$

which satisfy the Gaussian relation, but are as well close to the exact ones, Eq. (15). The agreement is reduced if one uses in Eq. (20) the simplified form without the correction factor, i.e. $F(\omega_1, \omega_2) = G(\omega_1)G(\omega_2)$. The latter approximation yields $\langle \omega^2 \rangle = 8t^2$, $\langle \omega^4 \rangle = 152t^4$.

We also note that $D(\omega = 0) = 1.367\ t$ within the RPA, in comparison to $1.447\ t$, obtained by the moment method, Eq. (16).

## III. NUMERICAL METHOD AND TESTS

An efficient numerical method which allows the study of $T > 0$ static and dynamical quantities in small correlated systems, has been recently introduced by the present authors [15], and applied to the dynamical conductivity [3] and to the spin susceptibility [16] in the $t - J$ model. The method, here used for the evaluation of the current-current correlation function $C(\omega)$, is based on two essential ingredients:

(a) the calculation of approximate eigenfunctions $|\psi_k^n\rangle$ and $|\tilde{\psi}_k^n\rangle$, $k = 1, M$, with corresponding energies $\epsilon_{nk}$ and $\tilde{\epsilon}_{nk}$, generated by the Lanczos diagonalization procedure from the initial functions $|n\rangle$ and $j|n\rangle$, respectively, and

(b) on a reduction of the full trace to a partial random sampling over basis functions $|n\rangle$,

$$C(\omega) = Z^{-1} \pi e_0^2 \sum_n^{N_0} \sum_{m,k}^{M} e^{-\beta \epsilon_{nm}} \langle n|\psi_m^n\rangle \langle \psi_m^n|j|\tilde{\psi}_k^n\rangle \langle \tilde{\psi}_k^n|j|n\rangle \delta(\omega - \epsilon_{nm} + \tilde{\epsilon}_{nk}),$$

$$Z = \sum_n^{N_0} \sum_m^{M} e^{-\beta \epsilon_{nm}} |\langle n|\psi_m^n\rangle|^2. \qquad (22)$$

It is important to note that both $N_0$ and $M$ are much smaller than the full-basis dimension $N_{tot}$, hence the requirements for the $T > 0$ calculation are comparable to the $T = 0$ Lanczos method for response functions [9–11]. An important feature of the method is that it reproduces correct frequency moments of $\sigma(\omega)$ (for a finite system) at higher $T$ as well as a meaningful $T = 0$ result.

Some tests of the method, in particular in the connection with the problem of dynamical conductivity, have been presented already in Refs. [15,3]. When discussing the feasibility of the method two different aspects have to be considered: (a) the role of approximations within the chosen finite-size systems, and (b) the finite size effects due to the smallness of the systems studied.

(a) Approximations introduced by using $M, N_0 \ll N_{tot}$ seem to be quite well understood and controlled. At full sampling $N_0 = N_{tot}$ the number of Lanczos steps $M$ determines the number of correct frequency moments (at least for higher $T$) within the chosen system, or equivalently the frequency resolution in spectra to $\Delta\omega \sim \omega_{max}/M$, $\omega_{max}$ representing typically the energy span of the model. Since the information content in these moments is limited, in particular due to finite size effects, there is no point in using very large $M$, i.e. we restrict our calculation in most cases to $M < 240$. The effects of using reduced sampling $N_0 \ll N_{tot}$ are most pronounced at low $T$, where only symmetry sectors with the lowest energies contribute.

We test the performance of the method on the example of a system of 10 sites and $N_h = 3$ holes. While the full basis dimension is $N_{tot} = 15360$ in this case, we choose $M = 40$ and $N_0 = 180$ in our method, i.e. $M, N_0 \ll N_{tot}$. We compare results to the exact ones, calculated via the full diagonalization, which requires the diagonalization of the $420 \times 420$ matrix in the largest $S_z = 1/2, q = 0$ sector. In Fig. 3 we present $\sigma(\omega)$ spectra at different $T/t$, obtained by both methods (plotted with an additional broadening of $\eta = 0.05\ t$). It should be noted that the smaller sampling $N_0 = 180$ represents within the ground state sector only $N_1 = 45$ states. In spite of a small effective number of samples even the $T = 0$ result is (surprisingly) robust, indicating that the response of different samples is quite similar. It is also plausible that the effects of finite sampling are even less pronounced at higher $T$, as evident from Fig. 3.

(b) Turning to finite size effects an important conclusion can be drawn from Fig. 3. Even at $T > t$, spectra for the 10-site system exhibit nonuniversal features, which could be attributed to finite size effects. The latter entirely disappear in larger systems, e.g. for $N = 16$ (at same $T$) as presented in Sec.IV. The difference is clearly due to the dramatic increase in the number of states $N_{tot}$ with $N$, whereby the number of peaks in the spectra scales even as $N_{tot}^2$ (within our method as $N_0 M^2$). This leads to a large reduction of (finite-size induced) fluctuations in spectra at particular $\omega$, requiring very small additional broadenings $\eta \ll t, J$ to average them. Thus one of the crucial advantages of the new method is in its ability to study systems which yield very smooth, macroscopic-like dynamical spectra at $T > 0$.

The density of (many-body) states becomes very sparse on approaching the ground state. Consequently, the thermodynamic averaging becomes ineffective at $T < T^* \propto \overline{\Delta E_0}$, where $\overline{\Delta E_0}$ is an average level spacing in the low-energy sector. Generally, the crossover at $T^*$ can be monitored by the emergence of finite size gaps in spectra etc. For $\sigma(\omega)$ an additional test is the charge stiffness [19],

$$D_c = -\frac{1}{2N}\langle H_{kin,xx}\rangle - \frac{1}{\pi e_0^2}\int_{0+}^{\infty} \sigma(\omega)d\omega, \tag{23}$$

which measures the coherent charge propagation around the system. For a macroscopic system at $T > 0$ (with the current being a nonconserved quantity) one expects generally $D_c = 0$, although this does not appear to be the case for (nontrivial) integrable quantum many-body systems [20].

In Fig. 4 we present results for $D_c(T)$ for a single hole in the $t-$ ($J = 0$) model. This example is particularly simple, since the ground state is known to be ferromagnetic (FM) with $E_{kin} = -4t$ as the consequence of the Nagaoka theorem [21]. Hence at $T = 0$ only the coherent contribution at $\omega = 0$ remains in $\sigma(\omega)$, and $D_c(T = 0) = t/N$. Fig. 4 shows that $D_c(T)$ interpolates quite smoothly between $D_c = 0$ at $T = \infty$ and $D_c(T = 0)$. The crossover however becomes increasingly sharp at larger systems $N \geq 18$, where it is important to notice the qualitative improvement when doubling the size from $N = 10$ (studied mainly so far by the method of full diagonalization [22]) to $N = 20$. Even for the largest system $N = 20$ we find that for $T < 0.2t$ $\sigma(\omega)$ becomes dominated by the undamped coherent contribution. Therefore, in this regime low-$\omega$ spectra do not represent properly the thermodynamic behavior.

For $J = 0.3\,t$ the results for $D_c(T)$ are less regular. We show in Fig. 5 results for various $N_h > 1$ on a system of $N = 16$ sites. When judging the extent of deviations of $D_c$ from zero at $T > 0$, it is useful to compare values with the maximum possible ones, i.e. with the value of the sum rule at $T = 0$, $D_{max} = |E_{kin}(T = 0)|/4N$, as follows from Eq. (23). We notice from Fig. 5 that $D_c$ typically shows a rather abrupt transition from $D_c \sim 0$ (small deviations

could be attributed to finite sampling and to finite width of $\omega$-channels) to $D_c(T) \neq 0$ at the crossover temperature $T^*$, depending on the hole doping $n_h$ in the particular system. For $T < T^*$ the variation $D_c(T)$ can become quite unphysical, i.e. we get in some cases even $D < 0$. These phenomena are influenced by the particular use of boundary conditions and more sensible results can be obtained by the introduction of additional fluxes (or twisted boundary conditions) [19]. Nevertheless we are here interested only in the regime $T > T^*$. From Fig. 5 it follows quite clearly that $T^*$ is minimum, i.e. $T^* \sim 0.1\,t$, for the intermediate (optimum) doping regime $0.1 < n_h < 0.3$, and is larger for the underdoped as well as for the overdoped systems.

## IV. NUMERICAL RESULTS

### A. Single hole mobility

The properties of a single hole are the basis for the understanding of the conductivity in the low-doping regime. Let us first consider results for the prototype, but notrivial, problem of the dynamical diffusion $D(\omega)$, Eq. (14), at $J = 0$ and $T \gg t$. The spectrum, as presented in Fig. 6, has been obtained for $N = 20$ and $M = 200$ to achieve high frequency resolution. For comparison we show also the RPA result, calculated from Eqs. (19,20).

It should be noted that for $N = 20$ we reproduce already the exact frequency moments (for an infinite system) up to the fourth order, Eq. (15), i.e. $\langle \omega^2 \rangle = 3t^2$ and $\langle \omega^4 \rangle = 33t^4$. The overall agreement of the RPA result with the numerical one is reasonable, taking into account the approximations involved in the RPA. It is however quite evident from Fig. 6 that in contrast to the RPA the actual $D(\omega)$ is *nonanalytical at $\omega = 0$*. We reproduce the same phenomenon in $N = 18$ system, but not within the same systems with $J > 0$. We do not have a proper explanation for this effect. A possible interpretation can be given in terms of finite ferromagnetic-like clusters. Probability for larger FM regions is decaying fast, on the other hand holes exhibit a nondiffusive motion in these regions, so that the interplay of both effects could result in a nonanalytical $D(\omega \sim 0)$. Such behaviour is not uncommon in the transport phenomena, known e.g. in connection with long-time tails [23].

The spectra for $J = 0$, but finite $T < \infty$, are presented in Fig. 7(a). Plotted are dynamical mobilities $\mu(\omega)$, Eq. (14), again calculated for $N = 20$. Comparing Fig. 7(a) with Fig. 6 we conclude that for $T \geq t$ the spectra essentially retain the high-$T$ form, i.e. $\mu(\omega) \sim D(\omega, T = \infty)/T$. Spectra change qualitatively for $T < t$, where the quasiparticle central peak starts to emerge at low $\omega$, and the incoherent broad background (peaked at $\omega \sim t$) vanishes on approaching $T \to 0$. In this case the relevant quasiparticle at $T \ll t$ represents the FM polaron. However, our systems are too small to be able to obtain some more quantitative conclusions on the FM polaron relaxation.

$J > 0$ does not influence spectra essentially at $T > t$, since according to Eq. (15) $J < t$ only slightly increases the width $\langle \omega^2 \rangle$. As seen from Fig. 7(b), there are qualitative differences appearing for $T < J$. Namely, for $T \to 0$ we are dealing with a nontrivial ground state of the antiferromagnetic (AFM) polaron, which has an enhanced mass but a nonvanishing incoherent spectral component at $\omega > J$.

Finally we extract from dynamical quantitites the dc values $\mu_0 = \mu(\omega \to 0)$, or equivalently an effective diffusion constant $D_0 = T\mu_0/e_0$ (which is not equal to a proper diffusion constant for $T < t$). It is plausible that the dc values can be obtained in a unique way only for $T > T^*$ where $D_c \ll D_{max}$. In Fig. 8 we present numerical results for $D_0(T)$, both for $J = 0$ and $J = 0.3\ t$, whereas the RPA ($J = 0$) result is shown for comparison. It is plausible that at $T > t$ finite $J > 0$ slightly reduces $D_0$, as follows also from Eqs. (15,16) applied to $J > 0$. On the other hand, both numerical and the RPA result show a decreasing tendency of $D_0(T)$ with decreasing $T < t$. In the resistivity $\rho(T) \propto T/D_0(T)$ this would show up as the deviation from the linear $T$-variation, i.e. $\rho(T)$ would remain rather constant (in particular for $J = 0.3\ t$) in the intermediate regime $T^* < T < t$. At low temperatures $T < t$ the relative reduction of $D_0$ due to $J > 0$ is much larger, indicating that the scattering of the hole on AFM excitations (magnons) is more efficient.

## B. Conductivity at finite hole dopings

Results for a single hole $N_h = 1$ can be interpreted also as low-doping results for the (sheet) conductivity $\sigma(\omega)$ in finite systems. In this sense we presented in [3] spectra for $n_h = 1/18$, which are essentially equivalent to the results for $n_h = 1/20$ shown above. By increasing the doping, $\sigma(\omega)$ could reveal new features, even at $T > T^*$. It is convenient to discuss in this case the dimensionless quantity $\sigma \rho_0$, where $\rho_0 = \hbar/e_0^2 = 4.1 k\Omega$ is the universal 2D sheet resistance.

We first show in Figs. 9 optical conductivities at fixed $J = 0.3\ t$, but at various dopings $n_h = 2/16, 3/16, 8/16$. Related spectra for $n_h = 4/16$ have been displayed already in [3], being very similar to the $n_h = 3/16$ case. One feature common to all $n_h$ considered (including $N_h = 1$) is the non-Drude type behavior for $\omega > t$. This confirms the belief that the incoherent motion of holes (dominating the high-$\omega$ response), as e.g. described within the RPA [6], can remain a valid concept even at large hole doping $n_h \sim 0.5$.

Some qualitative differences appear at low $\omega$. Results for $N_h = 1$ indicate that for $T < T^* \sim 0.7\ J$ a quasiparticle peak and a related quasigap starts to emerge at low $\omega$, as seen in Fig. 7(b). Some remains of this phenomenon could persist even at $n_h = 2/16$ in Fig. 9(a). On the other hand, no such effect is observed in the regime of 'optimum' doping with $n_h = 3/16, 4/16$, where no additional low-$\omega$ scale can be resolved down to the lowest $T = T^* \sim 0.1\ t$. On increasing $n_h$ further, $T^*$ increases substantially, indicating that the scattering becomes again less effective and the mean free path (at fixed $T$) longer. Hence, we cannot efficiently investigate low-electron densities $n_e < 0.5$, where one could presumably expect the Landau-Fermi-liquid behavior at low frequencies.

We calculate also the corresponding spectra for $J = 0$. Two of them are presented in Figs. 10. It is plausible that for $T > J$ spectra are not significantly influenced by $J$. Also at high dopings $n_h > 0.4$, $J$ becomes irrelevant at arbitrary $T$ (provided that $J < t$). E.g., for $n_h = 8/16$ we do not observe any significant difference between the $J = 0.3\ t$ and the $J = 0$ spectra. There appear however some systematic changes at $T < J$ within the low doping regime $n_h < 0.2$. Central peaks at the origin in Figs. 10(a),(b) (at $T < 0.5t$) seem to be sharper and more distinguishable than the ones in Figs. 9(a),(b). One could possibly attribute this difference to the known tendency towards the FM (at least partially) polarized ground state at $J = 0$, persisting even at $n_h > 0$.

With respect to experiments on cuprates [1], the most challenging is the case of 'optimum' hole doping $n_h \sim 0.2$ within the AFM correlated spin background generated by $J > 0$. In this regime optical measurements reveal a very universal behavior of $\sigma(\omega)$. In the previous paper [3] we showed that our model results for $n_h = 4/16$ reproduce (assuming $J/t = 0.3$, $t = 0.4$ eV) quantitatively sheet conductivities in BISCCO, YBCO [24] and LSCO ($x = 0.2$) [25]. It is very instructive to analyse the complex $\sigma(\omega)$ in terms of the effective frequency dependent relaxation time $\tau(\omega)$ and the effective mass $m^*(\omega)$, introduced via the memory function $M$ [26],

$$\sigma(\omega) = \frac{ie_0^2 S}{\omega + M(\omega)}, \qquad S = -\langle H_{kin,xx}\rangle/N, \tag{24}$$

and

$$\frac{1}{\tau(\omega)} = \frac{M''(\omega)}{1 + M'(\omega)/\omega},$$

$$\frac{m^*(\omega)}{m_t} = \frac{2n_h t}{S}\left(1 + \frac{M'(\omega)}{\omega}\right), \tag{25}$$

where $m_t = 1/2ta_0^2$ is the bare band mass. Using Eq. (25) one can formally rewrite $\sigma(\omega)$ in the familiar Drude form

$$\sigma(\omega) = \frac{in_h e_0^2}{m^*(\omega)[\omega + i/\tau(\omega)]}. \tag{26}$$

Using Eq. (25) we can evaluate from known $\sigma(\omega)$ both $\tau$ and $m^*$. Results for $n_h = 3/16, J/t = 0.3$ are presented in Fig. 11. From Fig. 11(a) it follows that in the whole low-$\omega, T$ regime we can remarkably well describe the behavior of $\tau$ with a simple law

$$\tau^{-1} = 2\pi\lambda(\omega + \xi T), \qquad \lambda \sim 0.1, \quad \xi \sim 3. \tag{27}$$

This is just one of the forms proposed within the MFL theory [14]. It should be also noted that obtained $\lambda$ is very close to the experimental one $\lambda \sim 0.11$ [27,1].

Also $m^*/m_t$ in Fig. 11(b) qualitatively agrees with experimental findings [24,27,1]. For quantitative comparison we should note that by lowering $T$ to experimentally investigated ones we expect the increase of $m^*$ at low $\omega$. On the other hand, assuming realistic values $a_0 = 0.38$ nm and $t = 0.4$ eV [4] we should take into account that $m_t \sim 0.6 \; m_0$.

Finally, we display in Fig. 12 results for dc resistivities $\rho(T)$, as extracted from $\sigma(\omega \to 0)$ (mainly presented and discussed already in [3]). We note that for $T > t$ and $n_h < 0.5$ slopes are approximately $d\rho/dT \sim \zeta \rho_0 k_B / n_h t$ with $\zeta \sim 0.4$, in comparison to the RPA value $\zeta \sim 0.72$, as deduced from Fig. 6. It is remarkable that $\rho(T)$ results are only weakly influenced by $J$ down to $T \sim J$, whereby it is plausible that the effect of $J$ nearly vanishes on approaching the electron-like regime $n_e < 0.5$.

## V. CONCLUSIONS

Main conclusions of numerical and analytical investigations, as presented above and in the preceding paper [3], can be summarized as follows:

(a) The finite-$T$ method yields within small systems meaningful (i.e. without pronounced finite size effects) spectra $\sigma(\omega)$ for $T > T^*(n_h)$. For the conductivity problem it is plausible to identify $T^*$ with the temperature, where the transport mean free path $l_0$ reaches (exceeds) the size of the system, i.e. $l_0(T^*) \sim \sqrt{N}$. This fact represents an independent information on transport. Within the studied system of 16 sites it is evident that $T^*$ depends quite significantly on doping. It is lowest for the 'optimum' doping $n_h \sim 0.2$ where $T^* \sim 0.3 \; J$. In the low doping regime $n_h < 0.1$ we find $T^* \sim 0.7 \; J$, while in the overdoped cases with $n_h \sim 0.5$ (where $J$ is irrelevant) $T^* \sim 0.5 \; t$.

(b) The single-hole dynamical diffusivity $D(\omega)$ at large $T \gg t$, evaluated from the lowest frequency moments or within the RPA, appears to be close to the Gaussian form. Numerical results confirm general features of analytical approximations, but for $J = 0$ also indicate on a fundamental nonanalytical behavior of $D(\omega \to 0)$.

(c) The single hole dynamics becomes quite sensitive on $J$ at low $T < t$, i.e. the hole can behave either as FM polaron for $J \sim 0$, or as an AFM polaron for $J > 0$.

(d) At finite doping, beyond the low-doping regime $n_h > 0.15$, both $\sigma(\omega)$ and $\rho(T)$ do not reveal an essential difference between $J = 0$ and $J = 0.3\,t$, at least not for $T > T^*$.

(e) At 'optimum' doping $n_h \sim 0.2$ $\sigma(\omega)$ spectra show a very universal form. For this case we extract the phenomenological relaxation time $\tau$ and the effective mass $m^*$. The form of $\tau(\omega, T)$ is consistent with the MFL proposal and moreover agrees with experimental values.

(f) Results for $\sigma(\omega)$ and $\rho(T)$ agree well with experiments without any adjustable parameters [3] (both $J$ and $t$ are well fixed either by experiments or by other considerations). This confirms the belief that the strong correlation effects, as embodied e.g. within the $t-J$ model, are crucial and possibly sufficient for the understanding of normal state properties of the cuprates and related materials.

## ACKNOWLEDGMENTS

The authors would like to acknowledge very helpful discussions with T.M. Rice and A. Ramšak. This work was supported by the Ministry of Science and Technology of Slovenia.

## Figure captions

FIG. 1. Some processes with nonvanishing contributions to $j^{(1)}$. Initial configurations are represented as: empty sites by open dots, fermions with up spins by full dots, and fermions with down spins by crossed dots. Straight lines show hopping and wavy lines spin flip processes. Numbering represents the order of processes.

FIG. 2. Some processes with nonvanishing contributions to $j^{(2)}$.

FIG. 3. $\sigma(\omega)$ spectra for $N_h = 3$ holes on $N = 10$ sites obtained via the full exact diagonalization (full line), and by using our method with the sampling $N_0 = 180$ and $M = 40$ Lanczos steps (dashed line).

FIG. 4. Charge stiffness $D_c(T)$ for a single hole in the $t - (J = 0)$ model for planar systems of various sizes.

FIG. 5. Normalized charge stiffness $D_c/D_{max}$ for a various number of holes $N_h$ in the model with 16 sites and $J/t = 0.3$.

FIG. 6. Dynamical diffusion $D(\omega)$ for a single hole at $T = \infty$ and $J = 0$, as obtained by the numerical calculation on a 20-sites system (full line), and the RPA theory (dashed line).

FIG. 7. Dynamical mobility $\mu(\omega)$ for a single hole at various $T$ in the planar $t - J$ model with: (a) $J = 0$, and (b) $J = 0.3\ t$.

FIG. 8. Effective diffusion constant $D_0(T)$ for a single hole. Shown are numerical results obtained on a system of $N = 20$ sites for $J = 0$ (circles) and $J = 0.3\ t$ (squares). Dashed line represents the RPA result.

FIG. 9. $\sigma(\omega)$ spectra for $J = 0.3\ t$ and various $T$ at finite hole dopings: (a) $n_h = 2/16$, (b) $n_h = 3/16$, and (c) $n_h = 8/16$. The smoothing width here and in Fig. 10 is $\eta = 0.07\ t$.

FIG. 10. $\sigma(\omega)$ spectra for $J = 0$ and various $T$ at finite hole dopings: (a) $n_h = 2/16$ and (b) $n_h = 3/16$.

FIG. 11. (a) Effective inverse relaxation time $\tau(\omega)$, and (b) the mass enhancement $m^*(\omega)/m_t$, for various $T$ in the system with $n_h = 3/16$ and $J = 0.3\ t$.

FIG. 12. Resistivities $\rho(T)$ at different dopings $n_h$ for $J = 0.3\,t$ (full lines) and $J = 0$ (dashed lines).